\documentstyle[epsf,aps,eqsecnum]{revtex}

\begin{document}

\title{Crossover between ballistic and diffusive regime of the 
spin-conductance
and CPP-GMR in magnetic multilayered nanostructures}
\author{S. Sanvito\thanks{e-mail: sanvito@dera.gov.uk},}
\address{School of Physics and Chemistry, Lancaster University,
Lancaster, LA1 4YB, UK and \\
DERA, Electronics Sector,
Malvern, Worcs. WR14 3PS UK}
\author{C.J. Lambert\thanks{e-mail:c.lambert@lancaster.ac.uk},}
\address{School of Physics and Chemistry, Lancaster University,
Lancaster, LA1 4YB, UK}
\author{J.H. Jefferson,}
\address{DERA, Electronics Sector, Malvern,
Worcs. WR14 3PS, UK}
\date{\today}
\maketitle

\begin{abstract}
We analyze the interplay between disorder and band
structure in current perpendicular to the planes (CPP)
giant magnetoresistance (GMR). We consider 
finite magnetic multilayers attached to pure crystalline leads, described
by a tight-binding
simple cubic two-band model ($s$-$d$). Several models of disorder
are considered, including random on-site potentials, 
lattice distortions, impurities, vacancies, 
and cross-section fluctuations.
Magneto-transport properties are calculated
in the zero-temperature zero-bias limit, within the 
Landauer-B\"uttiker formalism. 
Using a very efficient numerical scattering technique, 
we are able to perform simulations, 
over large length scales, and
to investigate spin-transport in the ballistic, diffusive and
localized regimes, as well as the crossover between them.
The competition between disorder-induced mean free path reduction and 
disorder-induced 
spin asymmetry enhancement
of the conductance highlights several different regimes of GMR.
\end{abstract}

\vspace{0.3in}


%\newpage

\section{Introduction}

Spin filtering in transition metal magnetic multilayers, 
which arises when the 
magnetizations of adjacent layers switch from an anti-parallel (AP)
to a parallel (P) alignment, is fundamental to the occurrence of 
giant magnetoresistance (GMR) \cite{gmr1,gmr2}.
The resistance in the anti-aligned state
can be as much as 100\% higher than the resistance 
with parallel alignment, leading to magnetic
field sensors with sensitivity far beyond that of
conventional anisotropic magnetoresistance (AMR) devices.
In the most common experimental setup, the current flows in the 
plane of the layers (CIP), and the resistance is measured with
conventional multi-probe techniques. Measurements in which the current
flows perpendicular to the planes (CPP) are more delicate because 
the resistances involved are small.
Despite this feature
the use of superconducting contacts \cite{msu}, sophisticated
lithographic techniques to form multilayered pillar structures
\cite{phil}, and electrodeposition
\cite{el1,el2,el3}, makes such measurements possible and to date
a large amount of experimental data has been produced 
(for recent reviews see references \cite{bau,ans}).
In the CPP configuration an
electron propagates across the whole multilayered structure, while 
in the CIP configuration it can in principle
traverse the system without being scattered at a
ferromagnetic/normal metal interface.
This makes the CPP configuration
more effective at filtering the current and consequently CPP-GMR
is generally larger than its CIP counterpart.
In what follows we shall focus our attention 
solely on CPP GMR.

On the theoretical side two fundamentally different approaches have 
been used to describe
CPP GMR. The first assumes that all the
transport is diffusive and is based on the semi-classical
Boltzmann equation within the relaxation time approximation. This
model has been developed by Valert and Fert
\cite{fer1}, and has the great advantage
that the same formalism describes both CIP and CPP experiments. 
It identifies the characteristic 
lengths of the problem and can include the effects of disorder into the
definition of the spin $\sigma$ dependent mean free path $\lambda_{\sigma}$ and spin 
the diffusion length $l_{\mathrm sf}$. 
Moreover it can be extended to describe the temperature
dependence of GMR \cite{fert2}.
In the limit that the spin diffusion length is much larger 
than the layer thicknesses
(infinite spin diffusion length limit), this model
reduces to a classical two current resistor network, in which 
additional spin-dependent scattering
at the interfaces is considered. The resistor network model
has been used since the early days of CPP GMR by the Michigan State University
group \cite{msu2}, and describes most of the 
experimental data. 
The parameters of the model are the magnetic
(non magnetic) metal resistivity $\rho_{\mathrm M}^*$
($\rho_{\mathrm N}^*$), the spin asymmetry parameter 
$\beta$ introduced through the spin dependent resistivity of
the magnetic metal $\rho_{\uparrow(\downarrow)}=
2\rho_{\mathrm M}^*(1-(+)\beta)$, 
the
magnetic/normal metal interface resistance 
per unit area $r_{\mathrm b}^*$
and the interface scattering spin asymmetry $\gamma$
introduced through the spin-dependent interface resistance
per unit area $r_{\uparrow(\downarrow)}=2
r_{\mathrm b}^*(1-(+)\gamma)$.
A good fit of the parameters has been shown to be possible,
and the same value can fit reasonably well both the CIP and
the CPP data \cite{msu2}. The limitation of such a model is
that it neglects the band structure 
of the system, and 
all the parameters are phenomenological. 
An extension of the model to include band structure
has been made recently \cite{ger1,ger2},
implementing the above transport theory 
within the framework of density functional theory in the local spin
density approximation. In this calculation, the scattering due to
impurities is treated quantum mechanically, while transport is described
semi-classically using the Boltzmann equation.
Material dependent studies are possible, but the calculations
are very computationally expensive and it is not possible to
deal with disordered systems.

The second theoretical approach to CPP GMR is based on the
quantum theory of scattering. 
Schep, Kelly, and Bauer \cite{bau2,bau3} showed that 
band structure alone
could account for the large CPP GMR found in Co/Cu multilayers.
Their calculations are based on local density functional theory and
the Sharvin resistance of a small constriction
formed from a pure crystalline infinite superlattice is calculated. 
This approach is
completely {\it ab-initio} but can deal only with clean systems, and
the unit cell must be small. 
To perform {\it ab-initio} studies of phase coherent
transport in disordered multilayers, more efficient
numerical techniques are required.
Tight-Binding methods based on $spd$ Hamiltonians derived from
first principle calculations have been employed by several groups
\cite{oxo,mato,noi}. On the one hand they can describe quite
accurately the band structure of the transition metal multilayers, and
on the other their computer overheads are more modest with respect
to density functional calculations.
As such they are suitable for numerical descriptions 
of long multilayers attached to realistic pure
crystalline leads.
Nevertheless the study of 
disorder using $spd$ tight-binding models is not trivial, because
the large number of degrees of freedom necessary to reproduce an
accurate band structure leads quickly to unmanageably large matrices.
The only calculations carried out to date involve either infinite
superlattices in the diffusive regime \cite{oxo} where small
unit cells must be used, or finite superlattices in which
disorder is introduced without breaking translational symmetry
in the direction perpendicular to the current \cite{mato,noi1}.
In the latter case the system is an effective quasi 1D system, whereas
real multilayers are 3D systems with
roughness at the interfaces which breaks translational 
invariance.

The aim of the present paper is 
to study three dimensional
GMR multilayers and to investigate
the effect of the disorder-induced cross-over between ballistic and
diffusive transport. To address this problem we consider a reduced
tight-binding model with two degrees of freedom ($s$-$d$) per atomic site. We
use a technique already employed to describe pure crystalline 
structures \cite{noi} to compute the zero-bias zero-temperature conductance
in the framework of the Landauer-B\"uttiker approach \cite{but}. 
We have optimized the 
calculation such that it scales sub-linearly with
the multilayer length.
Several models of disorder are introduced in order to mimic
defects, impurities, vacancies and lattice imperfections.
In the case of multilayered nanowires \cite{el1,el2,el3}
where the phase breaking length is
comparable with the wire cross-section, we consider the effects of
rough boundaries and confinement.
We show that phase coherent transport
in disordered magnetic multilayers may give rise to
behavior not describable by the Boltzmann approach and
discuss the relevance of these ``non-diffusive'' effects
in several new experiments.

The paper is organized as follows: in section 2 we describe our 
implementation of a numerical scattering technique capable of 
handling large systems and performing efficient averages over
large ensembles.
We also discuss the $s$-$d$ model which is the minimal 
Hamiltonian capable of capturing inter-band scattering.
In section 3 we present the main
results of this paper
and discuss the effect of different sources of disorder
and finally we 
conclude in section 4.

%%%%%%%%%%%%%%%%%%%%%%%%%%%%%%%%%%%%%%%%%%%%%%%%
%%%%%%%%%%%%%%%%%%%%%%%%%%%%%%%%%%%%%%%%%%%%%%%%
%%%%%%%%%%%%%%%%%%%%%%%%%%%%%%%%%%%%%%%%%%%%%%%%

\section{An efficient numerical scattering technique 
and models of disorder}

\subsection{Numerical Technique and a two band Model}

The numerical technique used 
in the present calculation has been
outlined in reference \cite{noi}, and describes an 
arbitrarily long finite
multilayer attached to two crystalline semi-infinite leads.
The spin-dependent conductance $\Gamma^\sigma$ of such a structure
is computed by evaluating the Landauer-B\"uttiker formula
\cite{but}

\begin{equation}
\Gamma^\sigma=\frac{e^2}{h}T^\sigma=\frac{e^2}{h}\sum_{k_\parallel}^
{\mathrm BZ}T^\sigma({k_\parallel})\;{,}
\label{con}
\end{equation}

where $e$ is the electronic charge, $h$ Planck's constant, and 
$T^\sigma$ ($\sigma=\uparrow, \downarrow$) the total spin-dependent
transmission coefficient,
defined as $T^\sigma={\mathrm Tr}\;t^\sigma t^{\sigma\dagger}$ with $t^{\sigma}$
the transmission matrix of the system.
The second equality is valid in the case of translational invariance
and the sum is taken over the 2D Brillouin zone in the direction
orthogonal to the current.
As a matter of notation we use the symbol $\parallel$ to indicate
$k$-points in
the plane of the layers, and the symbol $\perp$ to indicate
the direction of the current.
We completely neglect processes leading to spin mixing effects
and the two spin currents are treated separately. 
To utilize this expression in the presence of disorder,
we consider a disordered wire of finite cross-section,
which is periodically repeated in the transverse direction.
In the diffusive
limit, this coincides with the infinite spin-diffusion length limit of the Valert and
Fert theory which is equivalent to a classical resistor model.
We define $\Gamma^\uparrow_{\mathrm P}$ 
($\Gamma^\downarrow_{\mathrm P}$)
to be the conductance of the majority (minority) spins in the parallel
alignment, and $\Gamma^{\uparrow\downarrow}_{\mathrm AP}$ to be the conductance 
for both spins in the anti-parallel alignment, which yields for the GMR ratio,

\begin{equation}
{\mathrm GMR}=\frac{\Gamma^\uparrow_{\mathrm P}+
\Gamma^\downarrow_{\mathrm P}-2\Gamma^{\uparrow\downarrow}_{\mathrm AP}}
{2\Gamma^{\uparrow\downarrow}_{\mathrm AP}}
\;{.}
\label{gmr}
\end{equation}

The Hamiltonian for the whole system can be written

\begin{equation}
H=H_{\mathrm L}+H_{\mathrm LM}+H_{\mathrm M}+H_{\mathrm MR}+H_{\mathrm R}
\;{,}
\label{ham}
\end{equation}

with $H_{\mathrm L}$ ($H_{\mathrm R}$) the Hamiltonian of the semi-infinite
left- (right-) -hand lead, $H_{\mathrm LM}$ ($H_{\mathrm MR}$)
the coupling matrix between the left (right) lead 
and the multilayer, and $H_{\mathrm M}$ the Hamiltonian of the multilayer.
The key point is that we can decouple the calculation 
of the scattering channels in the leads from the calculation of an effective
Hamiltonian describing the multilayer. 
Consider first the retarded Green's function of the two decoupled
semi-infinite leads

\begin{equation}
g(E)=(E-H_{\mathrm L}-H_{\mathrm R}+i0^+)^{-1}
\;{.}
\label{gf}
\end{equation}

If the surface of the leads contains $M$ atoms each described 
by $n$ degrees of freedom, the Green's function $g$ is a 
$(nM)\times(nM)$ matrix. Hence the surface Green's function $g^{\mathrm S}$
involving only degrees of freedom of the left and right lead
surfaces is a $(2nM)\times(2nM)$ matrix whose matrix elements 
$g^{\mathrm S}_{ij}$ coupling the two leads vanish.
The block-diagonal matrix
$g^{\mathrm S}$ can be computed by evaluating the semi-analytic
expression given in reference \cite{noi}. 
Such a semi-analytic expression is valid if
the Hamiltonian
describing a crystalline lead can be written in the following 
trigonal form

\begin{equation}
H=\left(
\begin{array}[h]{rrrrrrrr}
... & ... & ... & ... & ... & ... & ... & ... \\
... & H_0 & H_1 & 0 & ... & ...  & ... & ... \\
... & H_{-1} & H_0 & H_1 & 0 & ... & ... & ... \\
... & 0 & H_{-1} & H_0 & H_1 & 0 & ... & ... \\
... & 0 & 0 & H_{-1} & H_0 & H_1 & 0  & ... \\
... & ... & ... & ... & ... & ... & ... & ... \\
... & ... & ... & ... & ... & ... & ... & ... \\
\end{array}
\right){\;},
\label{hamtrig}
\end{equation}

where $H_0$ is an hermitian matrix describing the coupling within
a cross-sectional slice of the lead, and $H_1$
($H_{-1}=H_1^\dagger$) is the coupling matrix between adjacent slices.
Once $g^{\mathrm S}$ is computed, the total surface
Green's function $G^{\mathrm S}$ of the leads + multilayer can 
be calculated by solving Dyson's equation

\begin{equation}
G^{\mathrm S}(E)=[(g^{\mathrm S}(E))^{-1}-\tilde{H}_{\mathrm M}]^{-1}
{\;},
\label{dys}
\end{equation}

where $\tilde{H}_{\mathrm M}$ is the $(2nM)\times(2nM)$
coupling matrix between the 
surface orbitals of the leads, which
can be obtained by
recursive decimation of the Hamiltonian 
$H_{\mathrm LM}+H_{\mathrm M}+H_{\mathrm MR}$.
Finally for a given $G^{\mathrm S}$
the scattering matrix elements can be obtained using
a variant of the Fisher-Lee relations \cite{colin1}.
In order to highlight the efficiency of this approach
in the case of disorder multilayers, we briefly describe how
the decimation technique recursively eliminates all
the internal degrees of freedom of the multilayer to 
yield the reduced matrix $\tilde{H}_{\mathrm M}$
coupling surface states of the leads. Suppose the total 
number of degrees
of freedom of the Hamiltonian $H_{\mathrm LM}+H_{\mathrm M}+H_{\mathrm MR}$
is $N$. It is possible to eliminate the $i=1$ degree
of freedom by reducing the $N\times N$ total Hamiltonian to an
$(N-1)\times (N-1)$ matrix with elements

\begin{equation}
H_{ij}^{(1)}= H_{ij}+\frac{ H_{i1} H_{1j}}{E- H_{11}}\;{.}
\label{dec1}
\end{equation}

Repeating this procedure $l$ times, we obtain the 
decimated Hamiltonian at $l$-th order

\begin{equation}
H_{ij}^{(l)}=H_{ij}^{(l-1)}+\frac{H_{il}^{(l-1)}H_{lj}^{(l-1)}}{E-H_{ll}^{(l-1)}}
\;{,}
\label{dec2}
\end{equation}

and after $N-nM$ iterations we obtain the $(nM)\times(nM)$
effective Hamiltonian

\begin{equation}
\tilde{H}_{\mathrm{M}}(E)=\left(
\begin{array}[4]{rr}
\tilde{H}_{\rm{L}}^*(E) & \tilde{H}_{\rm{LR}}^*(E)\\
\tilde{H}_{\rm{RL}}^*(E) & \tilde{H}_{\rm{R}}^*(E)\\
\end{array}
\right){\;}.
\label{haccasup}
\end{equation}

In this expression the matrices
$\tilde{H}_{\rm{L}}^*(E)$ and $\tilde{H}_{\rm{R}}^*(E)$
describe the intra-surface couplings respectively in the 
left and right surfaces, and $\tilde{H}_{\rm{LR}}^*(E)$
and $\tilde{H}_{\rm{RL}}^*(E)$ describe the effective coupling 
between these surfaces.
From equations (\ref{dec1}) and (\ref{dec2}) it is clear that only
matrix elements coupled to
the eliminated degree of freedom are redefined. Hence the recursive technique
becomes very efficient in the case of short-range interactions,
particularly in the case of nearest neighbor tight-binding 
models. 
Consider now a disordered multilayer composed of alternating magnetic (M)
and non-magnetic (N) layers of thicknesses
$t_{\mathrm M}$ and $t_{\mathrm N}$ respectively. Suppose that the multilayer consists
of $\mu$ repeated (N/M/N/M) units, that we call double bilayers (DB).
Since we consider only short range interactions, it is possible to 
decimate the Hamiltonian of the whole multilayer by building up
the following intermediate Hamiltonian 

\begin{equation}
H_{\mathrm M}=\left(
\begin{array}[h]{rrrrrrrr}
... & ... & ... & ... & ... & ... & ... & ... \\
V_0^\dagger & H_{\mathrm Li} & H_{\mathrm LRi} & 0 & ... & ...  & ... & ... \\
... & H_{\mathrm RLi} & H_{\mathrm Ri} & V_0 & 0 & ... & ... & ... \\
... & 0 & V_0^\dagger & H_{\mathrm L(i+1)} & H_{\mathrm LR(i+1)} & 0 & ... & ... \\
... & 0 & 0 & H_{\mathrm RL(i+1)} & H_{\mathrm R(i+1)} & V_0 & 0  & ... \\
... & ... & ... & ... & ... & ... & ... & ... \\
... & ... & ... & ... & ... & ... & ... & ... \\
\end{array}
\right){\;},
\label{mult}
\end{equation}

where $H_{\mathrm Li}$ ($H_{\mathrm Ri}$) describe the coupling within
the left (right) hand surfaces of the $i$-th cell (N/M/N/M) (i=1,..,$\mu$), 
$H_{\mathrm LRi}$ ($H_{\mathrm RLi}=H_{\mathrm LRi}^\dagger$)
describe the coupling between the left and right surfaces
of the $i$-th cell, and $V_0$ is 
the ``bare'' coupling between the first right-hand atomic plane
of the $i$-th cell and the first left-hand atomic plane of the
$(i+1)$-th cell, which is assumed to be the same for every cell
(this last condition is easily satisfied if the
first left-hand and the last
right-hand atomic plane of every (N/M/N/M) cell is disorder-free).
Equation (\ref{mult}) suggests a very convenient
implementation in which multilayers consisting of $\mu$ (N/M/N/M) cells
are built using the following procedure. Firstly we decimate 
a certain number $\nu$, of cells (N/M/N/M) in which disorder is introduced
everywhere except in the first and last atomic plane. Secondly the matrix
$H_{\mathrm M}$ of equation (\ref{mult}) is built, choosing randomly 
the order of the $\mu$ (N/M/N/M) cells.
Finally the matrix $H_{\mathrm M}$ 
is further decimated to yield $\tilde{H}_{\mathrm M}$. 
Note that $\mu^\nu$ possible different multilayers can be built
from a set of $\nu$ disordered unit cells,
and that the computation time scales as the number of (N/M/N/M)
cells and not as the total length of the scatterer.
This procedure can be further optimized, 
for instance by building $\nu^\prime$ new cells
(N/M/N/M)$\times 2$, and using these to 
form the multilayers. This turns out to be useful in
the case of very long samples.

\begin{figure}[h]
\begin{center}
\leavevmode
\epsfxsize=80mm \epsfbox{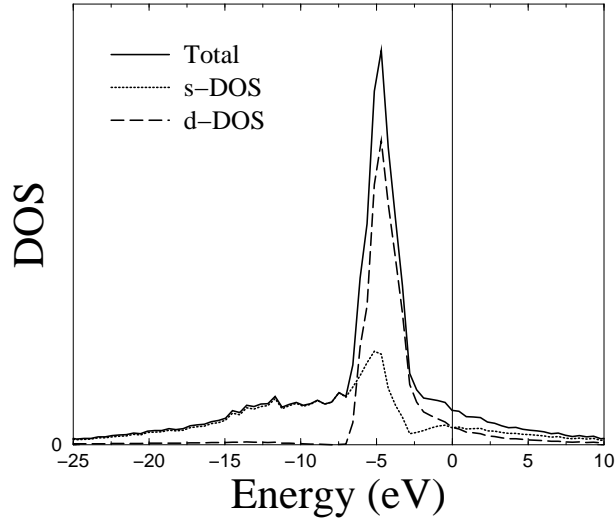}
\end{center}
\caption{ 
DOS obtained for the two-band model. The parameters used
are the ones corresponding to Cu in the Table \ref{tab1}. The
vertical line denotes the position of the Fermi energy
used in the calculation}
\label{pic1}
\end{figure}

\begin{figure}[h]
\begin{center}
\leavevmode
\epsfxsize=80mm \epsfbox{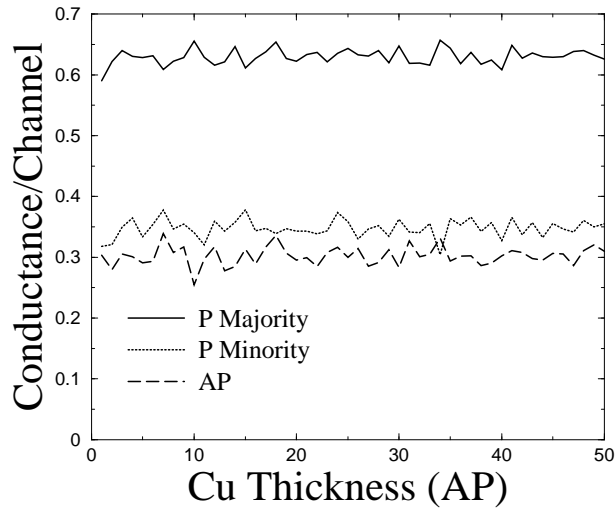}
\end{center}
\caption{
Conductances normalized to the number of open channels for
Co/Cu multilayers with Cu semi-infinite leads as a function
of the Cu layers thickness. This gives rise to a GMR of about 
60\%.}
\label{pic2}
\end{figure}

The technique for computing transport properties, 
is based on a three dimensional 
tight-binding model with nearest neighbor couplings on a simple 
cubic lattice and two degrees of freedom per atomic site. 
The general spin-dependent Hamiltonian is

\begin{equation}
H^\sigma=\sum_{i,\alpha}\epsilon^{\alpha\sigma}_{i}
c_{\alpha i}^{\sigma\dagger} c^\sigma_{\alpha i}
+\sum_{i,j,\alpha\beta}\gamma^{\alpha\beta\sigma}_{ij}
c_{\beta j}^{\sigma\dagger} c^\sigma_{\alpha i}
\;{,}
\label{spham}
\end{equation}

where $\alpha$ and $\beta$ label the two orbitals
(which for convenience we call $s$ and $d$),
$i,j$ denote the atomic sites and $\sigma$ the spin. 
$\epsilon^{\alpha\sigma}_{i}$ is the on-site energy which
can be written as 
$\epsilon^{\alpha}_{i}=\epsilon_0^{\alpha}+
\sigma h \delta_{\alpha {d}}$ with $h$ the
exchange energy and $\sigma=-1$ ($\sigma=+1$) for majority
(minority) spins. In equation (\ref{spham}),
$\gamma^{\alpha\beta\sigma}_{ij}=\gamma^{\alpha\beta}_{ij}$
is the hopping between the orbitals $\alpha$ and
$\beta$ at sites $i$ and $j$, and $c^\sigma_{\alpha i}$
($c_{\alpha i}^{\sigma\dagger}$) is the annihilation (creation)
operator for an electron at the atomic site $i$ in an orbital $\alpha$
with a spin $\sigma$. 
$h$ vanishes in the non-magnetic metal, and 
$\gamma^{\alpha\beta}_{ij}$ is zero if $i$ and $j$ do not
correspond to nearest neighbor sites.
Hybridization between the $s$ and $d$ orbitals is taken into
account by the non-vanishing term
$\gamma^{sd}$.
We have chosen to consider two orbitals per site in order to give an
appropriate description of the density of states of transition metals and
to take into account inter-band scattering
occurring at interfaces between different materials. The
DOS of a transition metal consists of narrow 
bands (mainly $d$-like) embedded in broader bands (mainly $sp$-like).
This feature can be reproduced in the two-band model, as shown
in figure \ref{pic1}.
The position of the Fermi energy with respect to the edge of 
the $d$ band determines the transport properties of pure
transition metals. For instance the current in silver is carried 
almost entirely by light effective mass $sp$ electrons with a small DOS, 
while in the minority band of Co or Ni it is carried by
heavy $d$ electrons with a large DOS. 
The hybridization at the Fermi
energy can also be important and for instance in copper the current
consists of an equal mixture of $sp$ and $d$ electrons.
In our earlier analysis of the material dependence of CPP GMR
\cite{noi} we identified large inter-band scattering as one of the main
sources of GMR. In particular we have shown that due to
inter-band scattering the conductance 
of a multilayer in the anti-parallel configuration
is always smaller than both spin conductances in the parallel 
configuration. 
It is possible to capture this feature
by choosing the parameters of the two-band model
to yield conductances 
as close as possible to those
obtained for the full $spd$ model \cite{noi}.
In the case of a heterojunction, the hopping parameters between different
materials are chosen to be the geometric mean of the hopping elements
of the bulk materials.
The parameters  for Cu and Co are presented in Table \ref{tab1}.
In figure \ref{pic2} we show the corresponding normalized 
conductance for
Co/Cu multilayers attached to semi-infinite Cu leads as a function
of the Cu layer thickness. 
We notice that as a consequence of inter-band scattering
the conductance
in the anti-parallel configuration is always the smallest,
a feature which is not present in a simple single-band model.
We believe this simple two-band model is the minimal model capable of describing in 
a semi-quantitative way the behavior of transition metals because it
includes the correct DOS and the possibility of scattering electrons
between high dispersion ($s$) and low dispersion bands ($d$).
 
\subsection{Models of Disorder}

\begin{figure}[h]
\begin{center}
\leavevmode
\epsfxsize=80mm \epsfbox{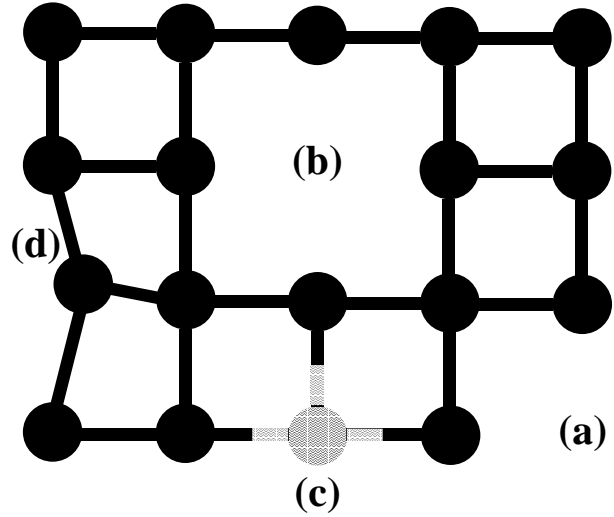}
\end{center}
\caption{
Schematic illustration of the disorder models considered: (a)
vacancy at the boundary of the cell (cross-section
fluctuation), (b) vacancy in the bulk, (c)
impurity (with hopping parameters the geometric mean of those for bulk
and the impurity),
(d) lattice distortion.}
\label{pic3}
\end{figure}

Figure \ref{pic3} shows the 
different models of disorder analyzed below.
The simplest model was introduced
by Anderson within the framework of the localization theory \cite{and}
and consists of adding a random potential $V$
to each on-site energy, with a uniform distribution of width
$W$ ($-W/2\leq V \leq W/2$), centered on $V=0$

\begin{equation}
\tilde{\epsilon}^{\alpha\sigma}_{i}=\epsilon^{\alpha\sigma}_{i}+V
\;{.}
\label{and}
\end{equation}

This generic model of disorder
can yield arbitrary mean free paths and significant
spin-asymmetry in the conductance.
To obtain a more realistic description of disorder 
we also consider the r\^ole of lattice distortions,
which are known to be present
at the interfaces between materials with different lattice constants.
Moreover in the case of electrodeposited nanowires, contamination by 
impurities
is unavoidable, and lattice distortions occur in the vicinity of such point 
defects. In what follows
we model lattice distortions by scaling the hopping
parameters between nearest neighbors. It has been proposed
\cite{ok} and confirmed numerically \cite{papa} that the following scaling
law for the tight-binding hopping $\gamma^{\alpha\beta}$ is valid

\begin{equation}
\gamma^{\alpha\beta}=\gamma^{\alpha\beta}_0\cdot (1+\delta r)^{-(1+\alpha+\beta)}
\;{,}
\label{scale}
\end{equation}

where $\gamma^{\alpha\beta}_0$ is the hopping element 
for atoms at the equilibrium
positions $r_0$, $\alpha$ and $\beta$ are the angular momenta of
the orbitals forming the bond, and $\delta r$ is the displacement
from the equilibrium position relative to $r_0$ ($\delta r=\Delta r/r_0$
with $\Delta r$ the displacement from the equilibrium position). 
Hence the $s$-$s$
hopping scales as $(1+\delta r)^{-1}$, the $d$-$d$ as
$(1+\delta r)^{-5}$ and the $s$-$d$ as $(1+\delta r)^{-3}$.
Note that it has been recently proved \cite{ger3} that in 3d transition
metals contaminated with 3d and 4sp impurities the variation
of the nearest neighbor distance in the proximity of an impurity
never exceeds $\sim 5\%$, which is within the limit of validity
of equation (\ref{scale}). In the following we will consider
uniform distributions of lattice displacements with zero mean.

As mentioned above,
in electrodeposited GMR nanowires, because of the dual-bath deposition
technique, the magnetic layers are contaminated
by non-magnetic impurities up to 15\% in concentration 
\cite{ans3}, while a negligible concentration of magnetic impurity atoms
is present in the non-magnetic layers.
To describe this feature we have introduced non-magnetic impurities in
the magnetic layers of the multilayer. An impurity is modeled by
substituting a magnetic ion by a non-magnetic ion
(ie Cu instead of Co for the materials considered)
at an atomic site. The on-site energy of the 
impurity is assumed to be the same as the bulk material
forming the impurity (ie bulk Cu for Cu impurities), and the
hopping tight-binding parameters depend on the
type of sites surrounding the impurity. We do not introduce
correlation between impurities and hence there are no clustering effects.
Although this model is quite primitive and does not take into account
perturbations of atoms in the proximity of the impurity,
density functional calculations \cite{ger4}
have shown that a good estimate of the resistivity of
transition metal alloys in the low concentration limit is
possible by considering only perturbations of the first
nearest neighbors of the impurity. This suggests that our simple
models should give a correct qualitative description of a 3d impurity
in 3d transition metals.

As a third source of disorder we have considered the possibility of vacancies. 
A vacancy is introduced simply by setting an  
on-site energy to a large number, with all the 
hoppings to nearest neighbors set to zero. We do not consider
aggregation of vacancies and assume
a uniform distribution across the whole multilayer. 
Finally we model cross-section fluctuations of
GMR nanowires by examining a wire of finite cross-section
which is not repeated periodically in the transverse direction and
mimic 
the fluctuations along the wire by introducing vacancies
in the first monolayer at the wire surface.

In all the calculations with disorder, we consider finite cross-sections 
involving $5\times 5$ atomic sites, which we repeat
periodically using up to 100 $k_\parallel$-points
in the 2D Brillouin zone. In the case of cross-section fluctuations 
we compute the ensemble-averaged conductance of wires with 
finite cross-sections as large as $15 \times 15$ atomic sites. 
It is important to note that in sputtered or MBE multilayers
the typical cross-sections vary between $1\mu$m$^2$ and 1mm$^2$,
which is several times larger than the typical phase breaking 
length $l_{\mathrm ph}$. On the other hand in the case of electrodeposited 
nanowires the diameter of the wires is usually between
20nm and 90nm, but several wires are measured
at the same times thereby yielding the mean conductance
of an array of phase coherent nanowires, each with a cross-section
of the order of $l_{\mathrm ph}^2$. 

\section{Results and Discussion}

\subsection{Disorder-induced enhancement of the spin-conductance asymmetry}

\begin{figure}[h]
\begin{center}
\leavevmode
\epsfxsize=80mm \epsfbox{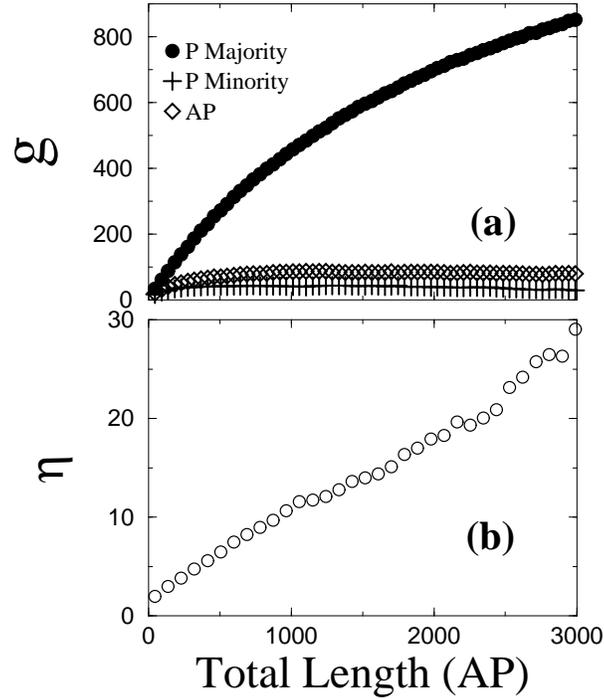}
\end{center}
\caption{Reduced conductance $g^\sigma$ and spin asymmetry
$\eta=g^\uparrow_{\mathrm P}/g^\downarrow_{\mathrm P}$
as a function of the multilayer length for Cu/Co multilayers
with random on-site potential. The random potential has a normal distribution 
of width 0.6eV, and the layer thicknesses are $t_{\mathrm Cu}=8$AP 
and $t_{\mathrm Co}=15$AP. Each point corresponds to a cell
Co/Cu/Co/Cu of total thickness 46AP.}
\label{pic4}
\end{figure}

In this section we consider effects produced by Anderson-type
disorder, impurities and lattice distortions. 
Despite the fact that the disorder in each of these cases
is spin-independent the effect on transport is spin-dependent. 
In order to investigate the different conductance regimes that may occur and
their dependence on the magnetic state of the system it is convenient to
consider as a scaling quantity the average spin conductance $<\Gamma^\sigma>$
multiplied by the total multilayer length $L$ and divided by the number of
open scattering channels in the leads.
We define the resulting ``reduced'' conductance $g$ by means of the equation

\begin{equation}
g^\sigma=\frac{h}{e^2}\frac{<\Gamma^\sigma>}{N_{\mathrm{open}}}\cdot L
\;{,}
\label{smgi}
\end{equation}

where the number of open channels in the leads $N_{\mathrm{open}}$ in
the case of a finite system is proportional to the multilayer cross-section.
In the ballistic
limit $g$ increases linearly with a coefficient proportional to
the conductance per unit area, in the diffusive (metallic) limit 
$g$ is constant, and in the localized regime $g$ decays as
$g\propto \exp(-L/\xi)$ with $\xi$ the localization length \cite{been,kram}.
Consider first the case of a random on-site potential.
For Co/Cu multilayers with a width of disorder $W=0.6$eV, figure \ref{pic4}
shows the quantity $g$ in units of $e^2/h$
for the two spin sub-bands in the
P and AP configurations along with the ratio
$\eta=g^\uparrow_{\mathrm P}/g^\downarrow_{\mathrm P}$.
These results were obtained for a cross-section of $5\times5$
atoms, and layer thicknesses of
$t_{\mathrm Cu}=8$ atomic planes (AP) and $t_{\mathrm Co}=15$AP.
In figure \ref{pic4} the standard deviation of the mean
is negligible on the scale of the symbols, and each point corresponds
to an additional Cu/Co double bilayer.
From the figure it is immediately clear that the spin-asymmetry of 
$g$ (ie of the conductance) is increased by the 
disorder, which as a consequence of the band structure,
turns out to be more effective in the minority
band and in the AP configuration.
In fact
the disorder has the effect of spreading the DOS beyond the
band edge, but does affect the centre of the band. The relevant quantity is
the disorder strength defined as the ratio $r_\alpha$ between the width
of the distribution of random potentials and the band width
$r_\alpha=W/\gamma_\alpha$. For the set of parameters that we have chosen 
the disorder
strength of the $s$ and $d$ band is respectively $r_d=0.7$ and $r_s=0.22$.
Since the current in the majority band of the P configuration
is carried mostly by $s$ electrons, for which the disorder strength
is weak, the majority spin sub-band will 
not be strongly affected by the disorder. 
In contrast in the minority band and in both bands in the AP
configuration, the current is carried by $d$ electrons, for which
the scattering due to disorder is strong.
 
A second remarkable
result is that in the P configuration
the almost ballistic majority electrons can co-exist with 
diffusive minority carriers. In the
regime of phase coherent transport the definition
of spin-dependent mean free paths for individual materials 
within the multilayer is not meaningful, and one must
consider the spin-dependent mean free path for the whole multilayered structure.
Hence we introduce the elastic mean free path for the majority
(minority) spin sub-band in the P configuration 
$\lambda^\uparrow_{\mathrm P}$ ($\lambda^\downarrow_{\mathrm P}$) and for
both spins in the AP configuration $\lambda^{\uparrow\downarrow}_{\mathrm AP}$.
This is defined as the length at which 
the corresponding conductance curve $g(L)$ changes from linear to constant 
(ie the length $L^*$ corresponding to the crossing point between the curve
$g(L)$ and the tangent to $g$ in the region where $g$
is constant).
For the calculation in figure \ref{pic4} we estimate
$\lambda^\uparrow_{\mathrm P}>3000$AP, $\lambda^\downarrow_{\mathrm P}\sim 500$AP
and $\lambda^{\uparrow\downarrow}_{\mathrm AP}\sim 1000$AP.
All of these results are obtained at zero temperature and voltage.
At finite temperature,  
when the phase
breaking length $l_{\mathrm ph}$ is shorter than the elastic mean 
free path, $l_{\mathrm ph}$ 
becomes
the length scale of the system. It is possible to generalize
the Landauer-B\"uttiker approach to the transport in presence of
finite phase breaking length \cite{but2}. In this case the system
can be considered as a series of phase coherent scatterers of
length $l_{\mathrm ph}$, added in series through reservoirs that 
make the phases random, and the scattering properties of such a structure
are solely determined by elastic transport up to a length $l_{\mathrm ph}$.
If the loss of coherence occurs on length
scales longer than the individual layer thicknesses
$t_{\mathrm Co}$ and $t_{\mathrm Cu}$, 
the standard resistor approach is not valid, and aggregate of cells as long
as $l_{\mathrm ph}$ are the appropriate quantities to add in series.

\begin{figure}[h]
\begin{center}
\leavevmode
\epsfxsize=80mm \epsfbox{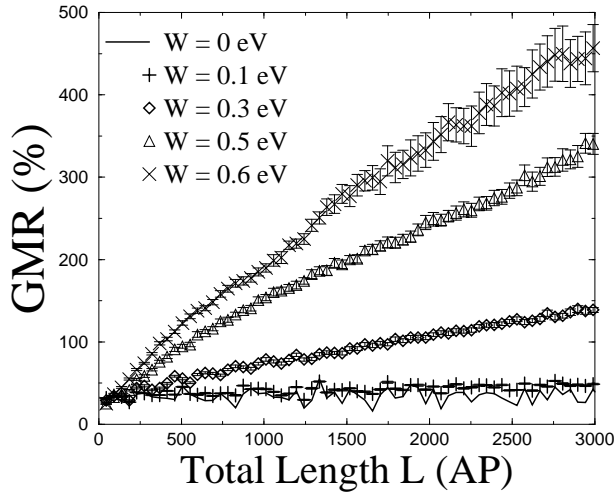}
\end{center}
\caption{GMR as a function of the total multilayer length for different
values of the on-site random potential. The layer thicknesses are 
$t_{\mathrm Cu}=8$AP 
and $t_{\mathrm Co}=15$AP and each point corresponds to a cell
Co/Cu/Co/Cu of total thickness 46AP.}
\label{pic5}
\end{figure}

\begin{figure}[h]
\begin{center}
\leavevmode
\epsfxsize=80mm \epsfbox{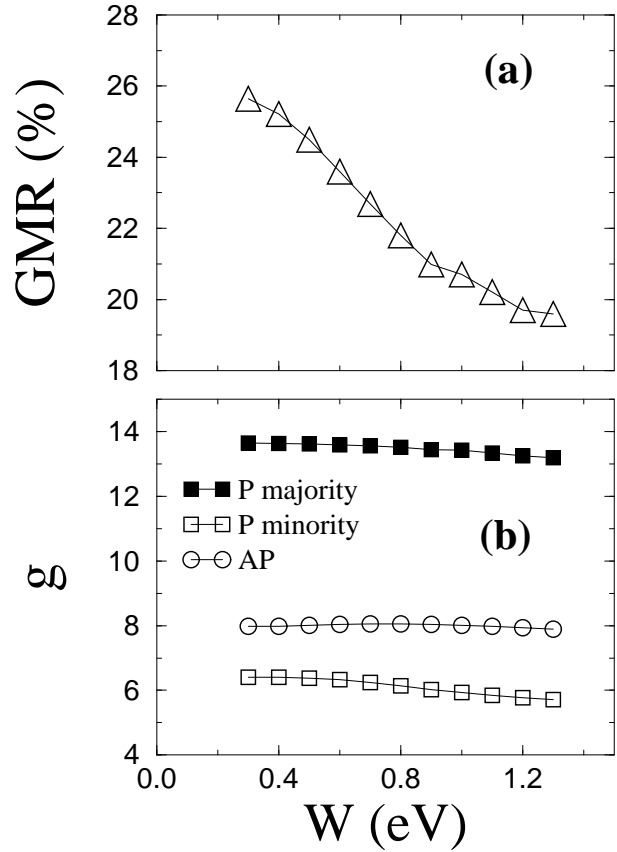}
\end{center}
\caption{GMR and reduced spin conductances as a function of the width $W$ 
of the normal distribution
of on-site random potentials for a single
Co/Cu/Co/Cu cell with Co and Cu thicknesses of 5AP.}
\label{pic6}
\end{figure}

Turning now our attention to GMR, it is clear from figure
\ref{pic4} and equation (\ref{gmr}) that enhanced spin asymmetry 
will increase the GMR ratio because of the
high transmission in the majority band. In figure \ref{pic5}
we present the GMR ratio as a function of the total multilayer length
for different values of the width of the distribution of the
random potential. From the figure we conclude that GMR
strongly increases as a function of the disorder strength and that this
is due to the increasing of the spin polarization of the conductance.
We also notice that the standard deviation of the mean GMR
increases as a function of disorder and of 
the multilayer length. This is due to the approaching of 
the AP conductance to the localized regime, in
which the fluctuations are expected to be large. The results of figures
\ref{pic5} seem to be in contradiction with the published
results of Tsymbal and Pettifor \cite{oxo}. In that case an analogous kind
of disorder was employed together with an accurate
$spd$ tight-binding model, and the GMR ratio turned out to decrease with
increasing disorder. They calculated the conductance
for an infinite diffusive system using a small disordered unit 
cell in the direction of the current, namely a Co$_4$/Cu$_4$ cell
(the subscripts indicate the number of atomic planes).
To check this apparent contradiction we have calculated the conductances
and the GMR ratio for a Co$_5$/Cu$_5$/Co$_5$/Cu$_5$ unit cell attached
to pure crystalline Cu leads. Apart from the resistances of the interfaces
with the leads, the conductance for this system is proportional to the
conductance calculated in ref \cite{oxo} and
figure \ref{pic6} shows that the GMR ratio for such a short system
does indeed decrease with disorder strength.
This shows that for small cells, when the mean free path is much longer 
than the cell itself, the increase of all the resistances is not fully 
compensated by an increase of their spin-asymmetry, and this
gives rise to a decrease of GMR. In contrast for thicker layers,
provided the transport
remains phase coherent, asymmetry builds up with increasing $L$ and
the resulting GMR ratio increases.

\begin{figure}[h]
\begin{center}
\leavevmode
\epsfxsize=80mm \epsfbox{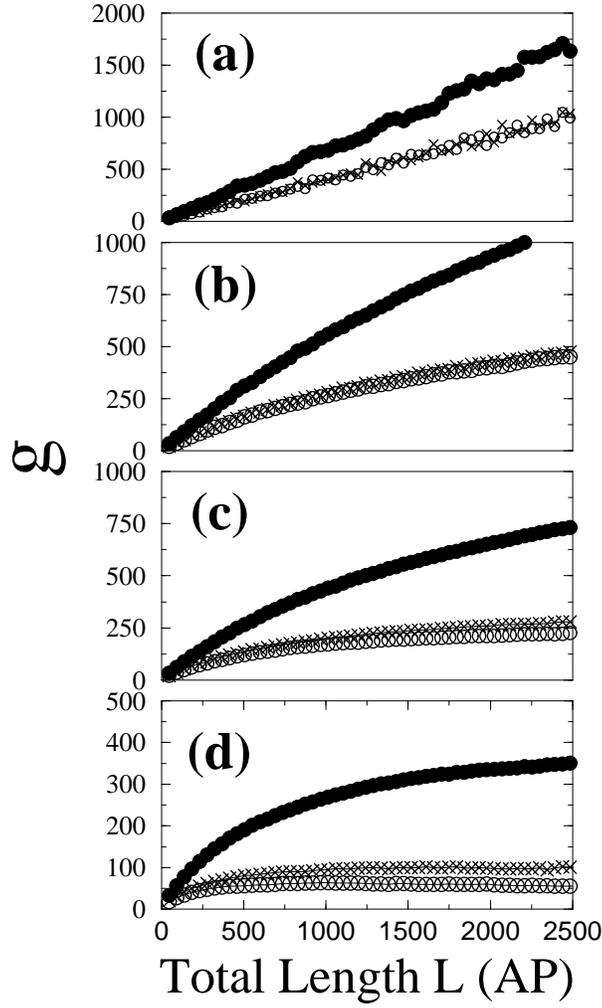}
\end{center}
\caption{Reduced spin conductances $g$ for different width of a normal
distribution of lattice distortion: (a) $\delta r=0$,
(b) $\delta r=0.02$\%, (c) $\delta r=0.03$\%, 
(d) $\delta r=0.05$\%. The symbols $\bullet$ ($\bigcirc$) 
represent the
majority (minority) spins in the P configuration, and $\times$
the AP configuration. The layer thicknesses are 
$t_{\mathrm Cu}=8$AP 
and $t_{\mathrm Co}=15$AP and each point corresponds to a cell
Co/Cu/Co/Cu of total thickness 46AP.
Note the different vertical scales
for the different disorders.}
\label{pic7}
\end{figure}

Consider now the effect produced by Cu impurities in the Co layers
and by lattice distortions.
The main features of both these kinds of disorder are very similar
to the case of a random on-site potential: the GMR ratio increases as
a function of disorder because of an increase in
spin-asymmetry. Again the quantity $g$
behaves quasi-ballistically for small lengths, followed by a
diffusive region and finally by a localized regime. The mean free path
at any disorder turns out to be longer for the majority spins
in the P configuration and the co-existence of ballistic majority electrons
with diffusive minority electrons is still possible.
This means that even in these cases spin-independent disorder
produces spin-dependent effects. Similar arguments to the one used
for the on-site random potential can be applied. 
In fact, in the case of impurities,
we note from Table \ref{tab1} that the alignment between
the majority band of Co and the conduction band of Cu is better than that
of the minority band of Co. 
Hence impurities are
less effective in the majority band than in the minority.
For lattice distortions, it is important to 
observe that the scaling of the hopping coefficients with the 
displacement from the equilibrium position is more severe for the $d$
orbitals (see equation (\ref{scale})).
Since the current in the majority band is $s$-like while in the minority
band and in the AP configuration it is $d$-like, this different
scaling will result in larger disorder-induced scattering
for the minority channel and for the
AP configuration. Figure \ref{pic7} shows
the reduced conductances $g$ for all the the spins in the case of 
uniform distributions of lattice displacements with different widths.
From the figure we can conclude that: i) the spin-conductance asymmetry
increases with increasing disorder ii) all the mean free paths decrease,
iii) the contrast between $g^\downarrow_{\mathrm P}$ and
$g^{\uparrow\downarrow}_{\mathrm P}$ increases with disorder.

\begin{figure}[h]
\begin{center}
\leavevmode
\epsfxsize=80mm \epsfbox{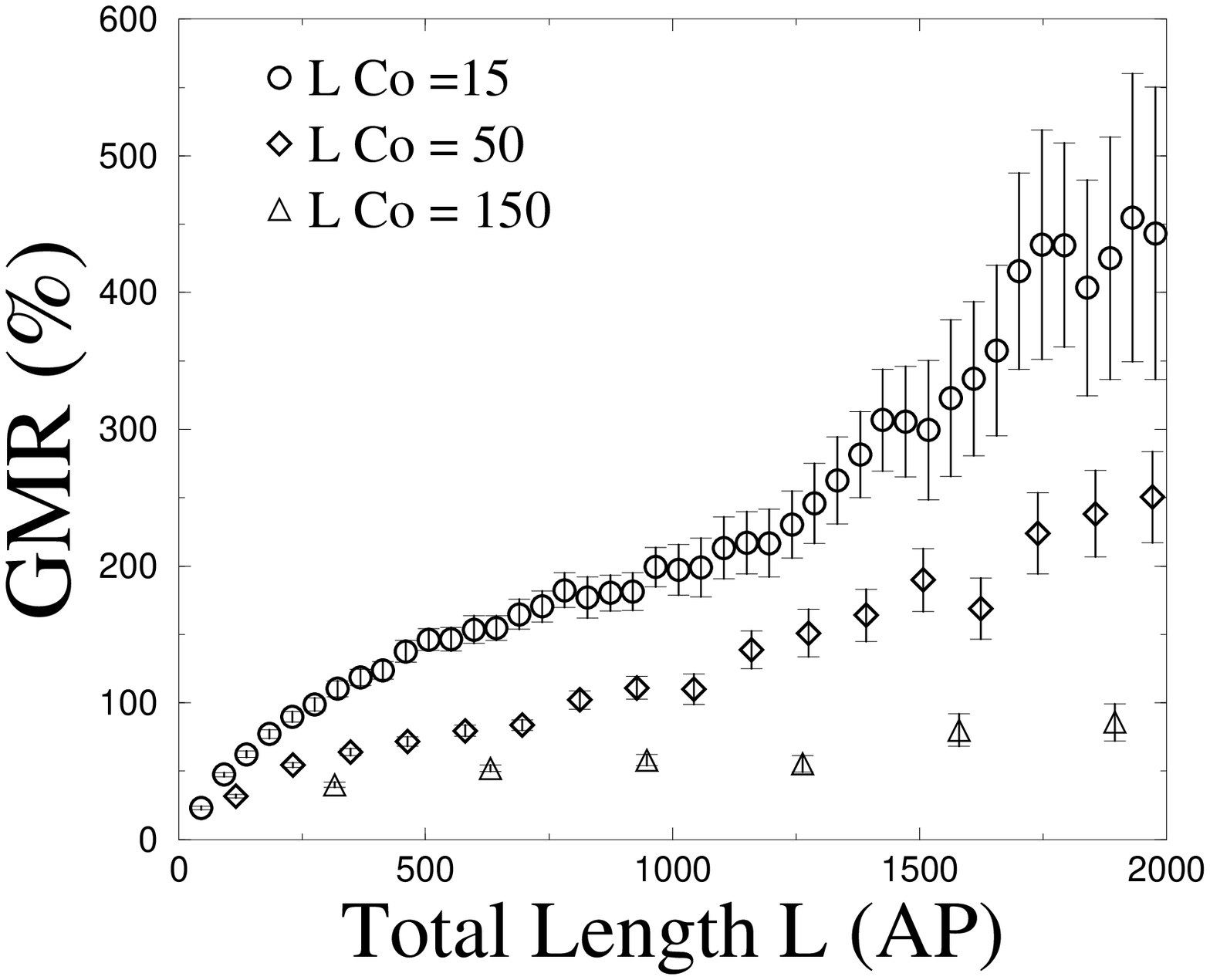}
\end{center}
\caption{GMR as a function of the number of double bilayers
for an impurity concentration of 8\%. The Cu thickness
is fixed to 8AP and the Co thickness is varied in order to show
the cross-over from a phase coherent regime
to a regime in which a resistor network model is
valid. Note that in the case of $t_{\mathrm Co}=$150AP the GMR
is almost independent of the total multilayer length.}
\label{pic8}
\end{figure}

We wish to conclude this paragraph with some final remarks about 
length scales involved. As mentioned above, since we are dealing
with phase coherent transport, the concept of mean free path within the
individual layers looses meaning, and we can only speak about the spin-dependent
mean free path of the whole multilayer (ie $\lambda^\uparrow_{\mathrm P}$,
$\lambda^\downarrow_{\mathrm P}$ and $\lambda^{\uparrow\downarrow}_{\mathrm AP}$). 
Nevertheless, if the mean free paths of both the spin sub-bands in the
P configuration 
extend over a length scale comparable with the cell Co/Cu
($\lambda^\uparrow_{\mathrm P}, \lambda^\downarrow_{\mathrm P} \sim 
t_{\mathrm Co}+t_{\mathrm Cu}$), the mean free path of the AP configuration
is simply given by

\begin{equation}
\lambda^{\uparrow\downarrow}_{\mathrm AP}=
\frac{\lambda^\uparrow_{\mathrm P}+\lambda^\downarrow_{\mathrm P}}{2}
\;{,}
\label{apmfp}
\end{equation}

and a
resistor network approach becomes valid.
We have checked this prediction by calculating the GMR ratio 
as a function of the number of
double bilayers for multilayers with different Co layer
thicknesses but the same concentration of impurities (8\%).
By increasing the Co thickness we can cross over from a regime in which
the resistor network is not valid at the scale of the bilayer thickness
to a regime in which the resistances of bilayers add in series.
In the first case we expect that the GMR ratio will increase as the 
number of bilayers
increases and in the second we expect a constant GMR.
The result for a Co thicknesses of respectively 150AP, 50AP and
15AP is presented in figure \ref{pic8}. 
Note that for a phase-coherent structure
the increase of GMR with the number
of bilayers is different from
the increase of GMR in diffusive systems when the total multilayer length
is kept constant (as predicted by the Boltzmann approach \cite{fer1}
and observed experimentally \cite{el2,el3,msu2}).
In the latter case the effect is due to an interplay between
the resistances of the different materials while in the former
it is due to an increase of the spin asymmetry of the current.
To date an increase of GMR with the number of bilayers
has been observed in the CIP configuration
\cite{chris}, while the same measurements in the CPP configuration
are still in progress \cite{chris2}.

\subsection{Reduction of mean free path}

\begin{figure}[h]
\begin{center}
\leavevmode
\epsfxsize=80mm \epsfbox{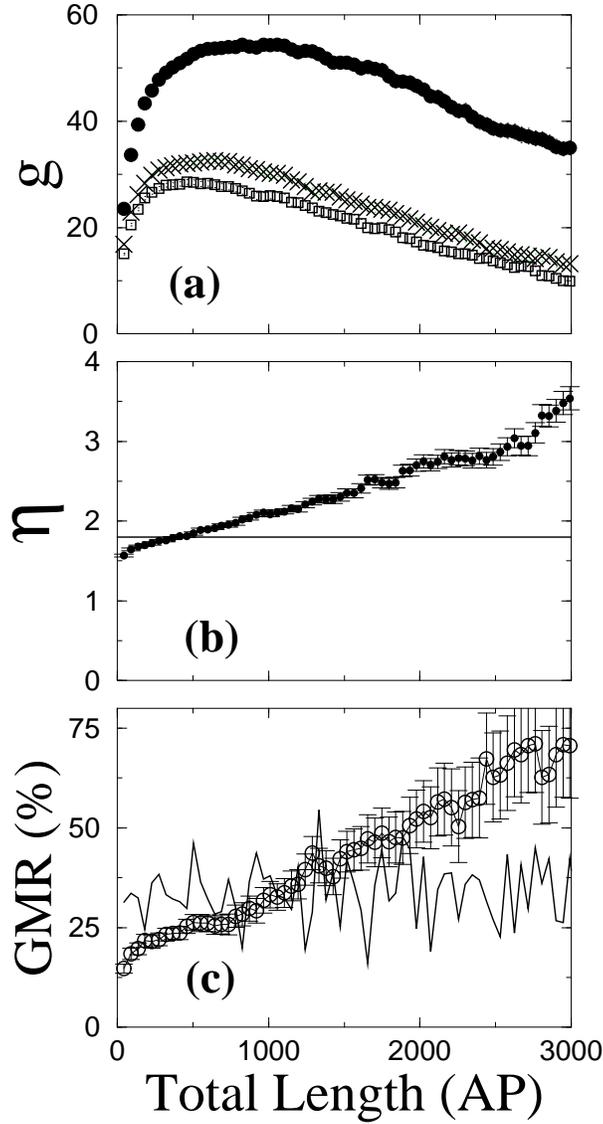}
\end{center}
\caption{Effects of vacancies on Co/Cu multilayers. Figure
(a) shows the reduced spin conductance for majority spin in the P
configuration ($\bullet$), for minority spins in the P
configuration ($\Box$), and for the AP configuration
($\times$). Figure (b) shows the spin asymmetry
of the conductance and figure (c) the GMR. 
The horizontal line of (b) represents the average spin
asymmetry of the conductance for the clean system. 
In figure (c) the symbols
$\bigcirc$ represent the system with vacancies and the 
solid line the disorder-free system. 
The vacancy concentration is 1\% and
the thicknesses are  $t_{\mathrm Cu}=$8AP and $t_{\mathrm Co}=$15AP.
}
\label{vac}
\end{figure}

In this section we consider the effect of vacancies and cross-section
fluctuations and their interplay with other sources of disorder
discussed in the previous section. 
We recall that cross-section fluctuations are modeled as
vacancies with a distribution concentrated at the boundaries of a 
finite cross-section multilayer. 
Hence we expect the qualitative behavior of vacancies
and cross-section fluctuations to be the same.
These sources of disorder do not act on the two spin sub-bands
in a selective way and produce only a small 
spin asymmetry. The main effect is to drastically reduce the 
elastic mean free paths of all the spins. In figure \ref{vac}
we present the reduced spin conductances $g^\sigma$, the 
spin asymmetry $\eta$ and the GMR ratio for a Co/Cu multilayer
($t_{\mathrm Cu}=8$AP, $t_{\mathrm Co}=15$AP)
with a vacancy concentration of 1\%. The results obtained
for cross-section fluctuations are very similar and are not
shown here.
Figure \ref{vac} shows that (in contrast with figure \ref{pic4}b)
the spin asymmetry
of the conductance is not greatly enhanced by the presence of vacancies.
For the parameters used in the present simulation 
$\eta$ varies from 1.6 to 3.5 for multilayers with a total
thickness ranging from 46 to 3000 atomic planes.
In contrast for the case of a random on-site potential of 0.6eV
figure \ref{pic4} shows that
$\eta$ varies from 2 to about 30 for the same range of multilayer lengths.
Moreover we notice that in the case of a random on-site potential
the spin asymmetry of the current is always larger than in the 
disorder-free case. In contrast, when vacancies are present,
the spin asymmetry of the current is smaller than the disorder-free
case for short multilayers
and becomes larger for longer multilayers. 
From figure \ref{vac} we can see that the cross-over length (that we denote
$l_{\mathrm cr}$),
defined as the length at which $\eta$ for a system with vacancies equalizes
$\eta$ for the disorder-free case, 
is comparable 
with the mean free path of the minority spins in the P configuration and of
the AP configuration. 
It is important to note that the reduction
of all the mean free paths with respect to the vacancy concentration
is very severe. The reduced spin conductances $g$ exhibit quasi-ballistic 
behavior for lengths up to $l_{\mathrm cr}$, and an almost localized
behavior for lengths larger than $l_{\mathrm cr}$. The diffusive
region is strongly suppressed and there is a small difference
between all the spin-dependent elastic mean free paths.
The spin asymmetry of the current can be enhanced by
increasing the vacancy concentration, but this produces
a further decreasing of the mean free paths and a further suppression
of the diffusive region, resulting in a global reduction of GMR
for lengths shorter than $l_{\mathrm cr}$.
For lengths longer than $l_{\mathrm cr}$ GMR is enhanced
and this is due to the approach of $g_{\mathrm AP}^{\uparrow\downarrow}$
to the localized regime.
To date there is no evidence of localization effects
in metallic magnetic multilayers and we believe that our results
are currently important only for lengths shorter than $l_{\mathrm cr}$.
To summarize, the main effects of vacancies are,
on the one hand to reduce the spin asymmetry of the current for lengths
shorter than $l_{\mathrm cr}$ and to enhance it for lengths larger than 
$l_{\mathrm cr}$, and on the other to reduce drastically the mean 
free paths for all the spins in both magnetic configurations. 
The cross-over length is comparable with the mean free path 
of the minority spin in the P configuration and GMR is
always reduced in the limit of quasi-ballistic transport.

The qualitative results obtained for vacancies are broadly 
mirrored by those of
cross-section fluctuations, although 
there are some differences.
The simulations with cross-section fluctuations have been 
carried out with a finite
cross-section, whereas for the case of vacancies where we have
considered a wire repeated periodically in the transverse
direction. 
When cross-section fluctuations are introduced, the disorder-induced
scattering 
scales as $P/S\propto 1/L$ with 
$P$ the perimeter, $S$ the 
area of the cross-section and $L=\sqrt{S}$. 
This introduces a new
length scale, namely the cross-section linear dimension $l_{\mathrm cs}=\sqrt{S}$.
If this length is shorter than the mean free paths, then
a reduction of GMR will take place for the same reasons 
as in the case of vacancies, whereas if the 
mean free paths are shorter than $l_{\mathrm cs}$, the effect of the
cross-section fluctuations will be weak and no further reduction
of the GMR will take place. Unfortunately, even with the optimized
technique presented in the previous section it is very difficult
to investigate the limit $\lambda\leq l_{\mathrm cs}$. We have
performed simulations with cross-sections up to 15$\times$15 atomic sites,
which is far below this limit, and have found no important deviations
from the case of vacancies. A cross-section of
15$\times$15 atomic sites corresponds to $P/S$ of $0.26 a_o^{-1}$
with $a_o$ the lattice constant.
This is comparable with the values of experiments 
\cite{el1,el2,el3} which we estimate range
between $0.005 a_o^{-1}$ and $0.025 a_o^{-1}$.
This suggests that the disorder strength in our simulations is
larger than experimental values and that the effects of the cross-section
fluctuations on GMR nanowires should be weak. On the other hand 
our model for cross-section fluctuations involves only
the first monolayer at the boundaries while in real systems
the roughness extends over several monolayers. Moreover long
range correlated surface roughness along the wires is likely
to be present in real systems because of the structure
of the nano-holes in which the wires are deposited. All these effects
may result in a drastic enhancement of the disorder strength due to surface
roughness and therefore a reduction of GMR.

\begin{figure}[h]
\begin{center}
\leavevmode
\epsfxsize=80mm \epsfbox{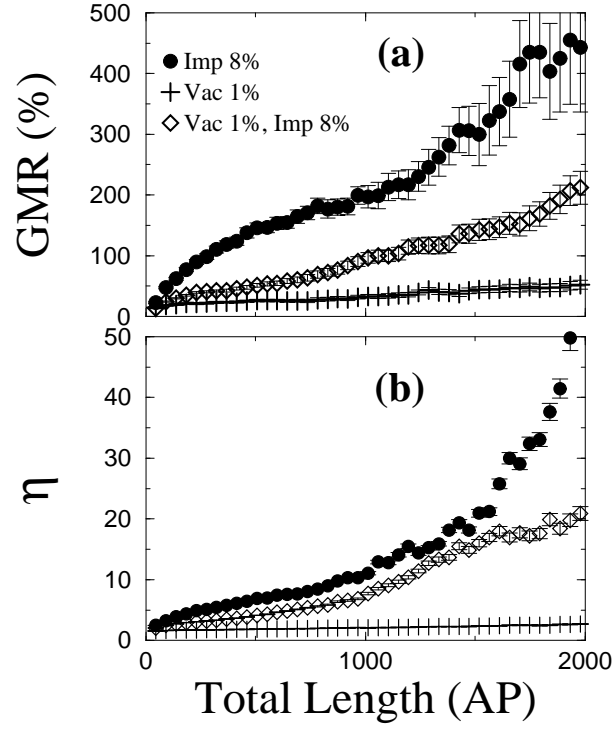}
\end{center}
\caption{Competition between vacancies and impurities.
Figure (a) shows the GMR ratio for Co/Cu
multilayers with only impurities ($\bullet$),
only vacancies ($+$) and impurities and
vacancies together ($\Diamond$).
Figures (b) shows the spin asymmetry of the current for
the same samples.
The layer thicknesses are 
$t_{\mathrm Cu}=$8AP and $t_{\mathrm Co}=$15AP.
}
\label{comp}
\end{figure}

A key result of the above simulation
is that the reduction of GMR due to
vacancies and cross-section-fluctuations may be compensated by a large
increase of the spin asymmetry of the conductance. To address this issue
we have performed simulations with both vacancies and non-magnetic impurities
in the magnetic layers. The GMR ratios and spin asymmetries of
the conductances are presented in figure \ref{comp} for Co/Cu
multilayers with different impurities and vacancies concentrations.
The figures shows very clearly that competing 
effects due to impurities and
vacancies can give rise to large values of GMR even for very
disordered systems. The same value of
GMR obtained in presence of impurities and vacancies can be obtained 
for a system
with only impurities, but at a lower concentration. The fundamental difference
between the two cases is that when impurities and vacancies
co-exist, all the mean free paths are very small and the large GMR is solely
due to the large spin asymmetry of the current. 

\section{Conclusions}

Due to the development of improved deposition techniques,
recent experiments \cite{chris2,msuord}
have revealed the need for a description of phase coherent transport,
which goes beyond the diffusive approach.
We have addressed this issue by extending a previously developed technique 
\cite{noi} to the case of disordered systems, where large ensemble averages 
are needed. We have presented several models of disorder within a 
two-band tight-binding model on a simple-cubic lattice. 
The model, despite its simplicity, can capture the relevant
aspects of transition metal multilayers and can provide a general
understanding of spin-dependent phase-coherent transport.
Moreover, because of the high efficiency of the technique, 
we have been able to investigate very long systems, different 
transport regimes (ballistic, diffusive and localized) and the cross-over 
between them.

We have shown that impurities, random on-site potentials and lattice 
distortions reduce the spin-dependent mean free paths, 
but at the same time increase the spin-asymmetry of the current and 
the GMR ratio.
In contrast, vacancies and cross-section fluctuations drastically
reduce all the spin-dependent mean free paths without largely increasing
the spin asymmetry of the current, and this produces a decrease of
the GMR, at least far away from the strong localized regime. Nevertheless
the effect of vacancies can be compensated by increasing the spin
asymmetry (for instance with impurities) and this can account for the
large GMR of electrodeposited nanowires.

\vspace{0.3in}

{\bf Acknowledgments}: 
This work is supported by the EPSRC, the EU
TMR Programme and the DERA.

\begin{table}[htbp]
\begin{center}
\begin{tabular}{|c|c|c|c|c|c|c|} \hline
\ {\bf Material} \ & \ \ \ $\epsilon_{\mathrm s}$ (eV) \ \ \ & \ \ \ $\epsilon_{\mathrm d}$ (eV)\ \ \ & 
\ \ \ ss$\sigma$ (eV) \ \ \ & \ \ \ dd$(\sigma,\pi,\delta)$ (eV)\ \ \ & \ \ \ sd$\sigma$ (eV)\ \ \ \ & 
 \ \ \ $h$ (eV)\ \ \ \\ \hline
{\bf Cu} & -7.8 & -4.0 & -2.7 & -0.85 & 1.1 & 0.0 \\  \hline
{\bf Co} & -4.6 & -2.0 & -2.7 & -0.85 & 0.9 & 1.6 \\  \hline
\end{tabular}
\caption{Parameters used in the calculations}
\label{tab1}
\end{center}
\end{table}

\end{document}